\documentclass[conference, 10pt]{IEEEtran}
\IEEEoverridecommandlockouts 
\usepackage[]{graphicx}
\usepackage{caption}
\usepackage{subfig} 
\usepackage{amsmath}
\usepackage{amssymb}
\usepackage{epsfig}
\usepackage{cite}
\usepackage{color}
\usepackage{balance}
\usepackage{amsmath}
\usepackage{amsfonts} 
\usepackage{graphicx}
\usepackage{pdfpages}
\usepackage[T1]{fontenc} 
\usepackage{array}
\usepackage{url}

\usepackage{color}
\usepackage{wrapfig}
\usepackage{lipsum}
\usepackage[hidelinks]{hyperref}
\usepackage{multirow}
\usepackage{tabularx}
\usepackage{tikz}
\usepackage{nth} 
\usepackage{algpseudocode} 
\usepackage{algorithm,algpseudocode}
\usepackage{breqn}

\begin{document}
\title{Predicting GPS-based PWV Measurements\\Using Exponential Smoothing}
\author{\IEEEauthorblockN{Shilpa Manandhar$^{\dagger}$\IEEEauthorrefmark{1},
Soumyabrata Dev$^{\dagger}$\IEEEauthorrefmark{3},
Yee Hui Lee\IEEEauthorrefmark{1}, and
Stefan Winkler\IEEEauthorrefmark{4}}
\IEEEauthorblockA{\IEEEauthorrefmark{1} School of Electrical and Electronic Engineering, Nanyang Technological University (NTU), Singapore}
\IEEEauthorblockA{\IEEEauthorrefmark{3} ADAPT SFI Research Centre, Trinity College Dublin, Ireland}
\IEEEauthorblockA{\IEEEauthorrefmark{4} School of Computing, National University of Singapore (NUS)}
\thanks{$^{\dagger}$ Authors contributed equally.}
\thanks{Send correspondence to Y.\ H.\ Lee, E-mail: EYHLee@ntu.edu.sg.}
\vspace{-0.6cm}
}

\maketitle

\begin{abstract}
Global Positioning System (GPS) derived precipitable water vapor (PWV) is extensively being used in atmospheric remote sensing for applications like rainfall prediction. Many applications require PWV values with good resolution and without any missing values. In this paper, we implement an exponential smoothing method to accurately predict the missing PWV values. The method shows good performance in terms of capturing the seasonal variability of PWV values. We report a root mean square error of 0.1~mm for a lead time of 15 minutes, using past data of 30 hours measured at 5-minute intervals.
\end{abstract}

\IEEEpeerreviewmaketitle

\section{Introduction}
Precipitable water vapor (PWV) values are an indicator of moisture content in the atmosphere and exhibit good correlation with rainfall events. There is a growing trend of using the PWV values derived from Global Positioning System (GPS) in detection and/or prediction of a rainfall event \cite{TGRS_RainNowcasting, shilpa_IGARSS_svm}. There are algorithms reported in the literature which predict rainfall events with lead times starting from $5$ minutes up to $6$ hours. The prediction window is affected by the resolution and availability of the GPS-PWV data. Generally, GPS-PWV values can be derived with a resolution of $5$ minutes~\cite{manandhar2018importance}. However, there are missing PWV values at certain hours. In this paper, we address this issue of missing PWV values, by proposing a method to predict the PWV values based on the past PWV values.

\section{GPS-based PWV Measurements}
\subsection{Computing PWV}
In this section, we briefly mention the methods to calculate PWV values from GPS measurements. The GPS signals are effected by two main delays in the troposphere layer of the atmosphere. They are Zenith Hydrostatic Delay ($ZHD$) and Zenith Wet Delay ($ZWD$). Out of these two delays, $ZWD$ is due to the water vapor content of the atmosphere. Therefore, PWV is derived from the $ZWD$ as shown by Eqs.\ (\ref{eq:PWV_ZWD},\ref{eq:PI}) \cite{shilpaPI}.
\begin{dmath}
    PWV=PI\cdot ZWD
    \label{eq:PWV_ZWD}
\end{dmath}
\begin{dmath}
	PI=[-\textrm{sgn}(L_{a}) \cdot 1.7\cdot 10^{-5} |L_{a}|^{h_{fac}}-0.0001] \cdot \cos\frac{2\pi(DoY-28)}{365.25}+0.165-1.7\cdot 10^{-5}|L_{a}|^{1.65}+f
    \label{eq:PI}
\end{dmath}
where $L_{a}$ is the latitude, $DoY$ is day-of-year, $h_{fac}=1.48$ for stations from the northern hemisphere and $1.25$ for the southern hemisphere. $f=-2.38\cdot 10^{-6}H$, where \textit{H} is the station height, which can be ignored for stations below $1000$m in altitude. In this paper, the $ZWD$ values are processed using the GIPSY OASIS software for a tropical IGS GPS station, ID: NTUS ($1.30$$^{\circ}$N, $103.68$$^{\circ}$E), with a temporal resolution of $5$ minutes.

\subsection{Predicting PWV Values}
Suppose $p_1$, $p_2$, \ldots , $p_t$ indicate the PWV values measured up to time $t$. We use triple exponential smoothing (TES) \cite{gardner1985exponential} to model the seasonal variations of the PWV values. The principal idea behind triple exponential smoothing is to apply exponential weights on the observations, with more weightage on recent observations. The TES weights are assigned on the level, trend and seasonal components of the time series. We use the $5$ minute interval PWV values in the form of time series data, and use only the historical PWV values to predict the future PWV values after time $t$. The future PWV values $p_{t+m}$ are modelled as:
\vspace{-0.2
cm}
\begin{dmath}
p_{t+m} =s_t+mb_t + c_{t-L+1+(m-1) \mod L},
\end{dmath}
\vspace{-0.4cm}
where $L$ is the length of a season, $s_t$ is the smoothed version of the constant part of observation, $b_t$ is the best estimate of the linear trend, and $c_t$ is the series of seasonal corrections. We benchmark our proposed method with two popular forecasting techniques -- persistence model and average model. The persistence model assumes that the forecasted PWV value remains constant as the latest PWV value, and is modelled as $p_{t+m} = p_t$. The average model works under the assumption that the future PWV values is the same as the average of the historical PWV values. It is modelled as $p_{t+m} = \frac{1}{t} \sum_{t}^{} p_t$.

\begin{figure*}[htb]
\centering
\includegraphics[height=0.22\textwidth]{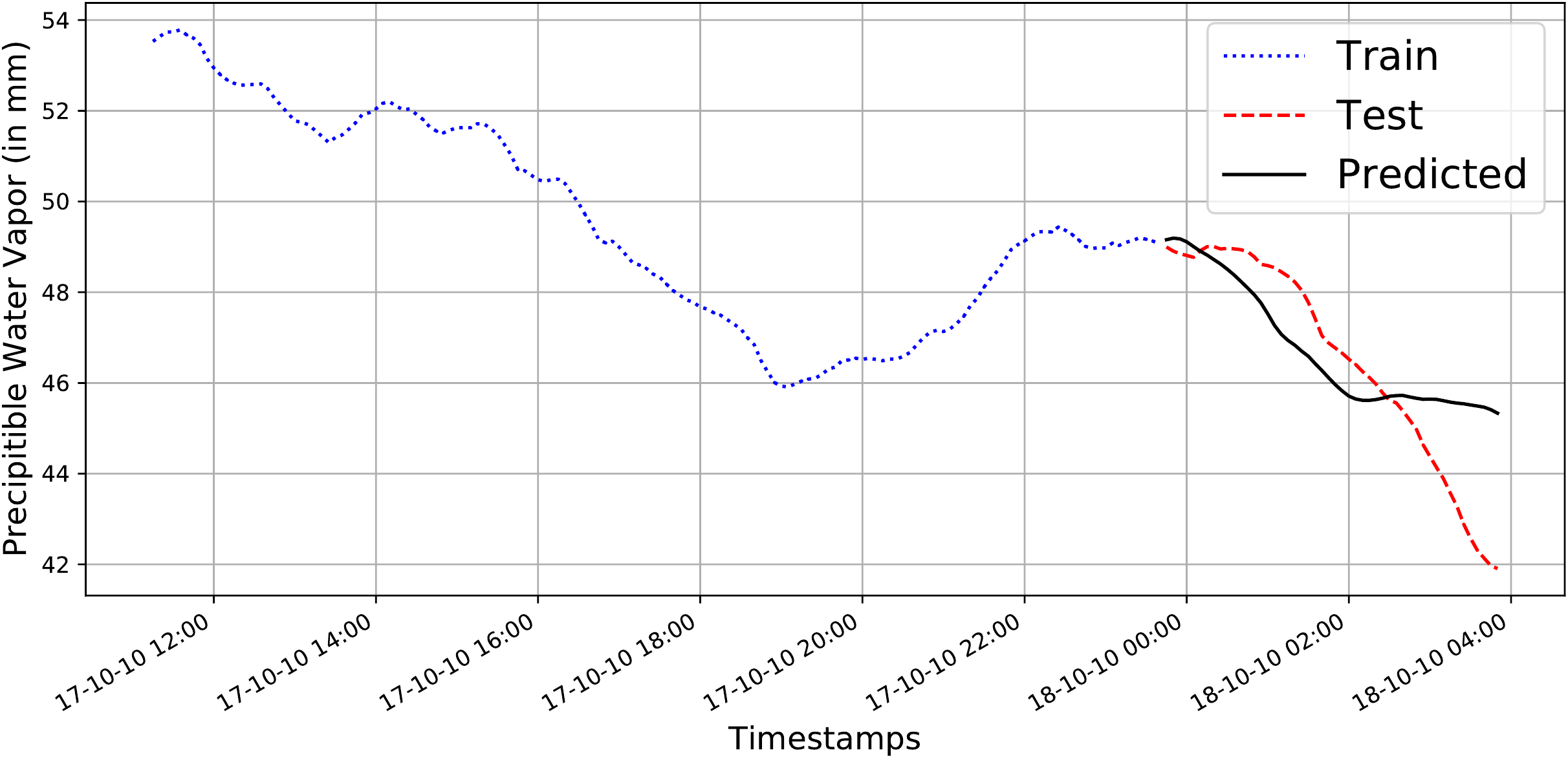}
\includegraphics[height=0.22\textwidth]{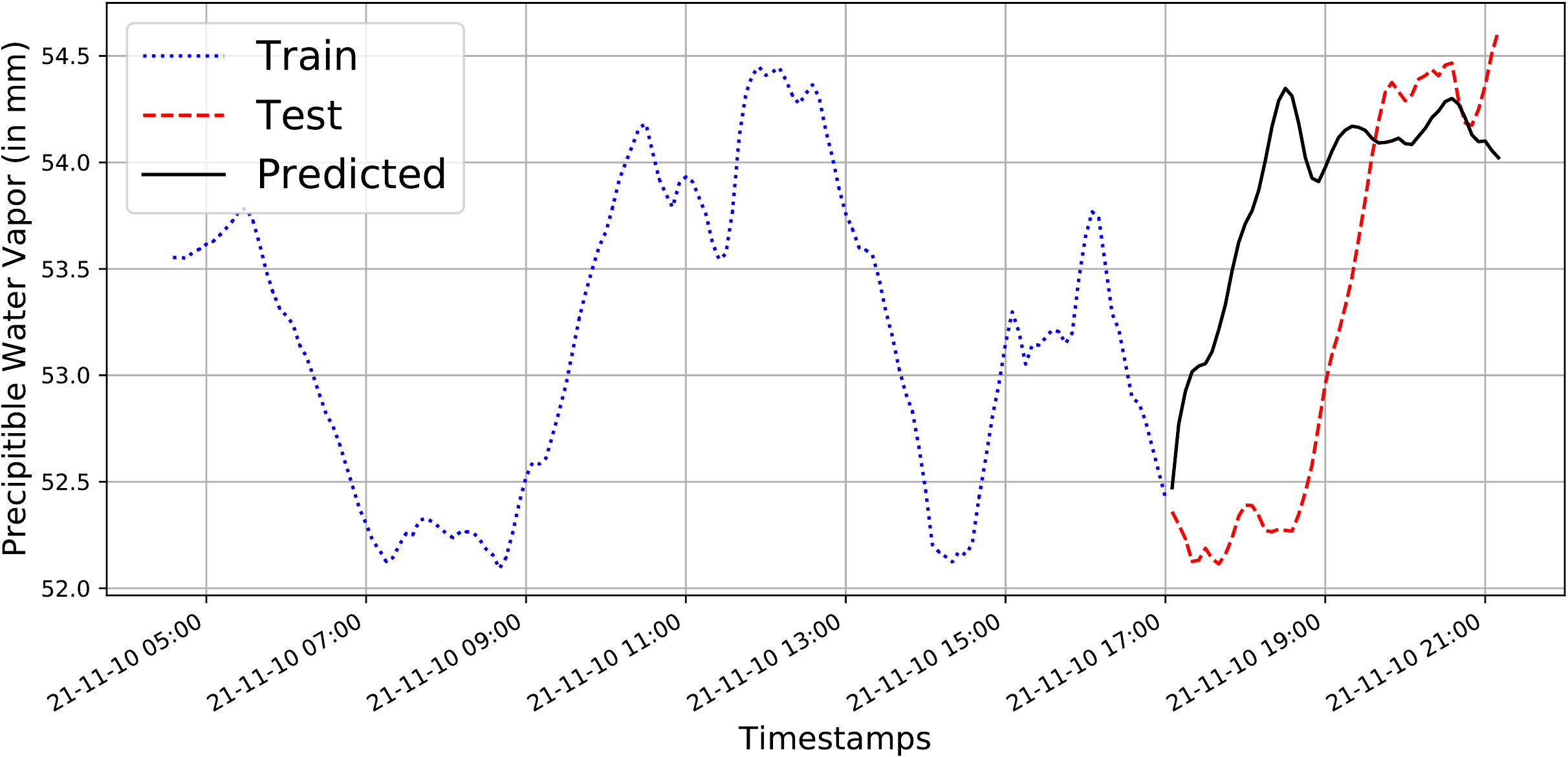}
\caption{Sample illustrations of the prediction of PWV values, along with the ground-truth values. We observe that the trend and seasonality of the future PWV values are captured with a good degree of accuracy.}
\label{fig:sample-examples}
\vspace{-0.4cm}
\end{figure*}

\section{Results \& Discussions}
In this section, we provide a detailed analysis~\footnote{The code of all simulations in this paper is available online at \url{https://github.com/Soumyabrata/predicting-pwv}.} of the forecasting of PWV values using exponential smoothing. The PWV values for the year $2010$ are computed for our chosen station ID NTUS. 

\subsection{Qualitative Evaluation}
Our proposed method can efficiently capture the seasonal variation of the PWV values, and provide a foundation for short- and long- term forecasting. Figure~\ref{fig:sample-examples} shows sample illustrations of the accuracy of PWV prediction. We use historical data of $30$ hours to predict the future PWV values. We observe that our proposed technique can capture the \emph{trend} of the future PWV values accurately. 

\subsection{Quantitative Evaluation}

\begin{figure}[htb]
\centering
\includegraphics[height=0.3\textwidth]{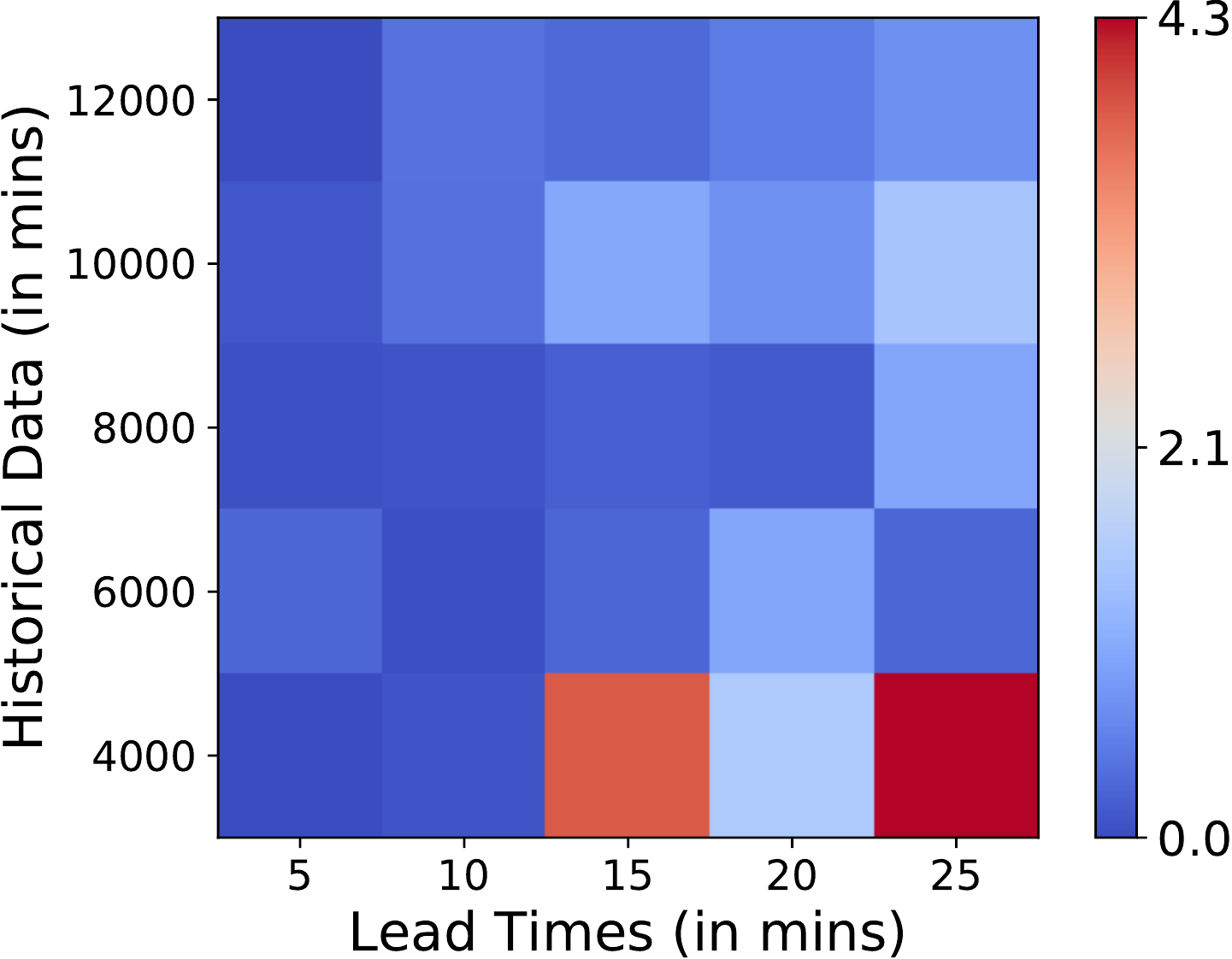}
\caption{Distribution of RMSE (mm) for a range of lead times and historical data.}
\label{fig:grid}
\vspace{-0.4cm}
\end{figure}

We use the Root Mean Square Error (RMSE) between the measured and predicted PWV values, in order to provide an objective evaluation of our proposed method. The performance of the prediction is dependent on two primary factors -- the amount of historical data that is considered for training the time series model, and the length of lead times to the forecast data. We use a varying range of historical data and lead times, in order to understand the impact of these variables on the forecasting performance. Figure~\ref{fig:grid} shows the impact of these two independent variables.

The corresponding error is color coded for a particular value of historical data and lead time. We repeat this experiment $10$ times for a chosen value of the two variables, in order to reduce any sampling bias. We observe that the error gradually increases with a lower value of historical data, and larger value of lead time. This makes sense as high amount of training data is required to model the seasonality properly, and the error accumulates as we predict higher lead times.

As a final comparison, we benchmark our proposed method with two popular baseline models, namely persistence and average. 
Table~\ref{table:compare} reports the average RMSE values (in mm) of the different methods in our dataset. We observe that the average model performs very poorly. Our proposed method shows a consistent improvement over the persistence model. This is due to the fact that PWV values remain fairly constant for shorter lead times. The forecasting performance can be further improved by incorporating other sensor data in addition to historical PWV values. 

\begin{table}[htb]
\centering
\normalsize
\caption{RMSE (mm) for varying lead times.}
\begin{tabular}{cccc}
\hline
Lead Time & Proposed & Persistence & Average \\
\hline 
~5 min                   & 0.061    & 0.086       & 10.433  \\
10  min                  & 0.078    & 0.144       & 9.525  \\
15 min                 & 0.101    & 0.259       & 7.028 \\
\hline
\end{tabular}
\label{table:compare}
\vspace{-0.6cm}
\end{table}

\section{Conclusion \& Future Work}
This paper applies an exponential smoothing method for predicting future PWV values using past PWV data. The exponential method shows better performance compared to the two other techniques. The RMSE values increase with longer lead time and less historical data. In future work, other meteorological data~\cite{dev2019multi} will be included for a better prediction of PWV values. 


\bibliographystyle{IEEEbib}

\end{document}